\documentstyle[11pt,aasms]{article}
\begin{document}

\lefthead{??}
\righthead{}

\title {{Spiral galaxies with {\tt WFPC2}: III. Nuclear Cusp Slopes\footnote
{Based on observations with the NASA/ESA Hubble Space Telescope,
obtained at the Space Telescope Science Institute, which is operated
by Association of Universities for Research in Astronomy, Inc.\
(AURA), under NASA contract NAS5-26555 }}}

\author{C.M.\ Carollo\footnote {Hubble Fellow}$^,$\footnote{Johns
Hopkins University, 3701 San Martin Dr., Baltimore, MD 21218
}$^,$\footnote{Space Telescope Science Institute, 3700 San Martin Dr.,
Baltimore, MD 21218} \& M. Stiavelli$^{4,}$\footnote {On assignment
from the Space Science Dept. of the European Space
Agency}$^,$\footnote {On leave from the Scuola Normale Superiore,
Piazza dei Cavalieri 7, I56126 Pisa, Italy } }

\begin{abstract}

In this paper, the third of a series dedicated to the investigation of
the nuclear properties of spiral galaxies, we have {\it (i)} modelled
the {\tt WFPC2} F606W nuclear surface brightness profiles of 41 spiral
galaxies presented in Carollo et al.\ 1997c, 1998 with the analytical
law introduced by Lauer et al.\ 1995, and {\it (ii)} deprojected these
surface brightness profiles and their analytical fits, so as to
estimate the nuclear stellar densities of bulges of spiral
galaxies. We find that the nuclear stellar cusps (quantified by the
average logarithmic slope of the surface brightness profiles within
0.1$''$-0.5$''$) are significantly different for $R^{1/4}$-law and
exponential bulges. The former have nuclear properties similar to
those of early-type galaxies, i.e.  similar values of nuclear cusps
for comparable luminosities, and increasingly steeper stellar cusps
with decreasing luminosity. By contrast, exponential bulges have
(underlying the light contribution from photometrically distinct,
central compact sources) comparative shallower stellar cusps, and
likely lower nuclear densities, than $R^{1/4}$-law bulges.

\end{abstract}


\section{Introduction}

The galactic nuclei are the repositories of low angular momentum
material sunk to the centers over the lifetime of the parent
systems. Therefore, they are likely to hold answers to important
questions related with the origin of structure in the parent
galaxies. In this perspective, establishing the demographics of
galactic nuclei along the entire Hubble sequence lies at the heart of
our understanding of the complex process of galaxy formation and
evolution.

Observations of nearby ellipticals and lenticulars with the Faint
Object Camera ({\tt FOC}), the Wide Field Planetary Camera ({\tt
WF/PC}) and the Wide Field Planetary Camera-2 ({\tt WFPC2}) aboard the
{\it Hubble Space Telescope} (HST) have revealed that the nuclei of
these galaxies are complex environments (e.g., Crane et al.\ 1993;
Jaffe et al.\ 1994; Lauer et al.\ 1995, hereafter L95; Forbes et al.\
1995; Carollo et al.\ 1997a, hereafter C97a; Carollo et al.\ 1997b,
hereafter C97b; Faber et al.\ 1997, hereafter F97).  They show surface
brightness profiles that increase down to the innermost point
measurable at HST resolution, i.e., $I(r) \propto r^{-\gamma}$ as $r
\rightarrow 0$ (where $I(r)$ is the surface brightness at the radius
$r$, and $\gamma > 0$); furthermore, several galaxies host stellar and
gaseous disks, unresolved nuclear spikes, double nuclei. These inner
features might possibly be related to the presence of massive black
holes (e.g., Lauer et al.\ 1996).

By contrast, much is still to be learned about the nuclear properties
of nearby spiral galaxies at HST resolution. F97 found that the surface
brightness profiles of the three Sa-Sb bulges present in their sample
show a behaviour similar to that of early-type spheroidals of
comparable luminosity. The same result was found by Phillips et al.\
(1996) for the three spirals of type earlier than Sc contained in
their {\tt WF/PC} F555W sample of 20 disk galaxies. Furthermore,
Phillips et al.\ found that later type spirals show instead (almost)
flat nuclear profiles, and suggested that the nuclear properties of
disk galaxies are more closely related to those of nucleated dwarf
galaxies than to those of elliptical galaxies. Further exploration is
necessary to assess how the nuclear properties scale with the
properties of the spheroidal component. This is likely to provide
feedback information about the epoch and processes of nucleus, bulge,
and, ultimately, galaxy formation.

In order to address these issues, we have performed a {\tt WFPC2}
snapshot survey in the F606W filter of the nuclei of 107 (mostly Sa to
Sc) disk galaxies. In paper I (Carollo et al.\ 1997c) and paper II
(Carollo et al.\ 1998) we have presented the 75 targets imaged so far
within our program. Our analysis shows that bulge-like structures are
present in most of the galaxies.  While in some cases these are
``classical'', smooth, featureless $R^{1/4}$-law bulges, in others
they are better fitted by an exponential profile (see also Courteau,
de Jong \& Broeils 1996, and references therein, for similar results).
The exponential bulges include two classes of objects: {\it (i)}
dwarf-looking systems, whose surface brightness profiles within
$\approx 15''$ are well fitted by a single exponential. These galaxies 
are strongly bulge-dominated; their surrounding, faint regions (``disk/halo'') 
show no signs of spiral arms, and  have typically a quiescent, i.e. non star 
forming, appearance.  {\it (ii)} Small exponential bulges embedded
in dominant (spiral-armed/star-forming) surrounding disks, i.e., the
inner exponential structures of double-exponential fits to the surface
brightness profiles within $\approx 15''$.  The exponential bulges
as-a-class are statistically fainter than the featurless, smooth
$R^{1/4}$ bulges, for constant disk luminosity and Hubble
type. Resolved, central compact sources are found in most of the
exponential bulges; the hosts of central compact sources often contain
a barred structure. The nature of these compact source, and in
particular their relation with e.g., star clusters and Seyfert 2
nuclei, is discussed in Carollo (1998).

In this paper, the third of the series, we investigate the relation
between the nuclear structure of spiral galaxies and the physical
properties of inner disks and/or bulges.  In particular, we {\it (i)}
present the results of modeling the nuclear surface brightness
profiles with the analytical law introduced by L95 (for the 43
galaxies of paper I and II for which we could perform the
measurements), {\it (ii)} deconvolve the surface brightness profiles
and their analytical fits in order to estimate the nuclear stellar
densities, {\it (iii)} study the nuclear properties as a function of
the global properties discussed in papers I and II (e.g., $R^{1/4}$
against exponential bulges), and {\it (iv)} compare the nuclear
properties of our sample with those observed in early-type
galaxies. The paper is organized as follows. In section 2 we briefly
summarize the properties of the sample, the data used in our
investigation, the procedure adopted for the data reduction, and the
steps performed to derive the surface brightness profiles. In section
3 we present the results of the analytical fits applied to the surface
brightness profiles, and of deconvolving data and models in spherical
symmetry. In section 4 we investigate the dependence of the nuclear
properties on global galactic properties. We conclude in section 5. 

\section{The Sample, the Data and the Data Reduction}

We address to papers I and II for details on the selection of the
total sample, the data, and the procedure adopted for their
reduction. Below we briefly summarize the important points.

We selected a total of 92 spiral galaxies (from the UGC catalog for
the northern hemisphere, Nilson 1973, and from the ESOLV catalog for
the southern hemisphere, Lauberts \& Valentijn 1989), which satisfy
the following constraints: {\it (i)} Angular diameter larger than 1
arcmin; {\it (ii)} Regular types {\it Sa, Sab, Sb,} and {\it Sbc};
{\it (iii)} Redshift $<$ 2500 km/s, to guarantee a high angular
resolution in physical size; {\it (iv)} An inclination angle,
estimated from the apparent axial ratio, smaller than 75 degrees.
Fifteen additional galaxies were chosen among E/S0 and S0/Sa, so as to
have a reference sample for comparison with the literature on
early-type galaxies.  In this paper we use the surface brightness
profiles of the 18 objects presented in paper 1, and of the 25 objects
presented in paper 2, for which the light distribution was regular
enough to allow us to perform the measurements. Some parameters for
the total sample of 43 galaxies studied in this paper are given in
Table 1. We adopt $H_o=65$ km/s/Mpc throughout this paper.

For each galaxy we acquired two {\tt WFPC2} F606W exposures of 400 and
200 seconds, respectively.  The observations were taken with (gain 15
and) the galaxy nucleus centered on the PC, which has a scale of
0.046$''$ px$^{-1}$, and a useful field of about $35''\times35''$. The
observations were carried out in fine lock. The data reduction
included: {\it (i)} the reprocessing of the raw images with the
standard {\tt WFPC2} pipeline in order to use the most recent
reference frames for flat fielding, bias and dark current subtraction;
{\it (ii)} the addition and cosmic-ray cleaning of the two images by
using the {\tt IRAF/STSDAS} task ``crrej''; {\it (iii)} the removal of
remaining hot pixels by interpolation; {\it (iv)} sky subtraction by
determining the sky values from the {\tt WFC} images, in areas
furthermost from the nucleus.  The derivation of the surface
photometry was carried out in IRAF. For the sample of paper I, we used
two different ellipse-fitting programs (``Galphot'', Franx,
Illingworth \& Heckman 1989, and ``Ellipse'', R. Jedrezjeski), and two
different approaches for the dust- and star-masking, so as to have an
estimate of the uncertainties introduced by the complex dust and star
forming structures.  Both approaches only correct for patchy dust
absorption and not for any smooth and extended dust component, which
remains undetectable in our data. For the sample of paper II, we only
used the task ``Ellipse'', since the previous study shows that the two
packages give equal results within the errors. Absolute photometric
calibration was obtained by applying the zero-point for the F606W
filter provided by Holtzman et al.\ (1995), i.e.
\begin{eqnarray}
mag_{F606W} &  = & -2.5  \log(counts/sec)+Z_{WFPC,F606W} \\
   & & +5 \log(0.046)+0.1\nonumber
\end{eqnarray}
with the zero-point $Z_{WFPC,F606W}$ equal to 22.084. The subsequent
term accounts for the pixel size. The constant shift of 0.1 mag was
added following Holtzmann et al.\ in order to correct for infinite
aperture. This calibration provides a rough conversion to $V$-Johnson
magnitudes (i.e., in average within $\approx 0.2$ magnitudes, given
the difference in bandpass between the Johnson $V$ and the F606W
filters, and the lack of a color term for our sample).  The calibrated
surface brightness profiles were corrected for Galactic extinction by
using the values published by Burstein \& Heiles (1984).

\section{Analytical Fits to the Nuclear Surface Brightness Profiles} 

One of the main purposes of any analytical description of the light
detected from astronomical sources is to provide a simple way to
quantify the most relevat features and their variations from object to
object within the same family of sources, and to compare these
features among objects of different families. For the early-type
galaxies, a popular description of the surface brightness profile has
been proposed by L95, and used by Byun et al.\ (1996) to describe the
nuclear properties of kinematically normal early-type galaxies, and by
C97a, C97b to describe those of elliptical galaxies with
kinematically-distinct cores.  It reads:

\begin{equation}
\label{physlaw}
I(r) = 2^{(\beta-\gamma)/\alpha} I_b (\frac {R_b}{r})^\gamma \Big[ 1 +
( \frac {r}{R_b})^\alpha \Big]^{(\gamma-\beta)/\alpha}.
\end{equation}

The parameter $R_b$ is the break radius at which, in giant elliptical
galaxies, the profile flattens to a more shallow -- but always
non-zero -- slope, $\gamma$ is the slope of the inner profile (i.e.,
$I(r)\sim r^{-\gamma}$ as $r\rightarrow0$), $\beta$ is the slope of
the outer profile, $\alpha$ controls the sharpness of the transition
between inner and outer profile, and $I_b$ is the surface brightness
at $R_b$.  The profile of equation (2) provides an adequate
description of the features present in the surface brightness profiles
of early-type galaxies along their entire luminosity sequence.

The structure that we detect in the nuclei of several disk galaxies
requires a clear definition of what are the relevant features in these
galaxies that we aim at quantifying. In particular, for the
exponential galaxies which host a perfectly centered compact source,
there is a conceptual choice which has to be made, namely whether to
include the central source in the analytical description of the galaxy
(in a few galaxies, the ``central'' compact source reported in Table 1
is actually slightly offset from the center of the outer
isophotes. Therefore, it has not been considered in the isophotal
fitting, and does not appear in the surface brightness profile).  In
principle, these systems could be described by an exponential profile
modified to include a steep nuclear cusp slope.  In order to check
whether this is a physically acceptable solution, we fitted the
exponential-bulge galaxies hosting a central compact source with a
light profile described by:
\begin{equation} 
I(R) = I_0 \left( 1+\frac{R_c}{R} \right)^\gamma \exp{(-R/R_s)}.  
\end{equation} 
For radii $R$ much smaller than the ``cusp'' radius $R_c$, this
profile describes a cusp with slope $\gamma$. $I_0$ is the central
brightness of the esponential component, and $R_s$ the esponential
scalelength.  No acceptable (i.e., physically meaningful) fit could be
found for any galaxy except ESO 499 G37.  This demonstrates that
``steep-cusp esponentials'' could indeed exist, and that our procedure
would be able to identify them. However, the fact that they generally
do not provide a good or meaningful fit supports the interpretation,
based on morphology, that central compact sources are photometrically
distinct components from their surroundings, and not simply a
steepening of the light profile.  On this basis, we considered these
sources as an {\it additional contribution} to the to-be-fitted,
underlying galaxy light. The radial profile of the latter is what can
be meaningfully compared to the radial profile measured for the
early-type systems. Indeed, the same choice was made by Byun et al.\
(1996), who excluded from their fits the light excess from the
resolved nuclei embedded in the bulges of a few galaxies of their
sample (e.g., NGC 1331, NGC 3599, NGC 4239). Therefore, in order to
perform a direct comparison with elliptical and lenticular galaxies,
we adopted equation (2) to describe the nuclear properties of spiral
galaxies.  Some details on the analytical ``L95 fits'' are given in
Appendix A.

There is an intrinsic uncertainty in disentangling the contribution of
the nuclear compact source from the {\it unknown} underlying galaxy
light.  In order to quantify this uncertainty, we performed several
fits to the same galaxy, by considering different inner radial cutoff
in its surface brightness profiles. The resulting inward
extrapolations of the best fits to the outer points may differ
significantly, but are likely to bracket the true shape of the
underlying galactic light, free from PSF-blurring and central compact
source contamination. We were able to fit 41 of the 43 galaxies (for
the two remaining objects, ESO 205G7 and NGC 3177, no acceptable fit
was obtained).  In Figure 1 we show the surface brightness profiles,
and the fits with the Point Spread Function (PSF) convolved models,
for the 41 galaxies. In Table 2 we list, for each galaxy, radial range
adopted for performing the fits, and the corresponding best fit values
for $R_b$, $\alpha$, $\beta$, $\gamma$ and $\mu_b=-2.5 \log I_b$. For
those objects for which changing the radial range for the fits lead to
significantly different solutions, we show in the table and in the
figure the two ``bracketing'' fits.

For each galaxy, the analytical fits to the surface brightness
profiles were used to derive the average logarithmic nuclear slopes
$\langle \gamma \rangle$ corrected for PSF and central compact source
effects. The average slopes provide a global representation of the
underlying projected galactic light within a defined radial
range. This was taken equal to 0.1$''$-0.5$''$ (similarly to what was
done by L95 and C97a for the early-type systems), which corresponds to
physical scales smaller than $\approx100$ pc at the average distance
of the galaxies ($\approx 25$ Mpc). In contrast to the values of the
individual parameters in the best fits, which for the same galaxy can
differ significantly from one best fit to the other (obtained within
different radial ranges), the $\langle \gamma \rangle$ values provide
a robust description of the nuclear surface brightness profiles
contributed by the underlying galactic light. For the galaxies for
which two fits are presented, we adopted the arithmetic mean of the
corresponding values of $\langle \gamma \rangle$. An estimate of the
uncertainty associated with these measurements is given by the
difference between the two values, and is typically smaller than $\sim
0.2$.  The values of the average logarithmic slopes inside
0.1$''$-$0.5''$, $\langle \gamma \rangle$, are also listed in Table 2.

\subsection{Deprojections}

The analytical fits to the surface brightness profiles were
deprojected in spherical symmetry by means of an Abel inversion
(Binney \& Tremaine 1987), to derive the luminosity density profiles
$\rho_{mod}(r)$. The same was done using directly the data points
[$\rho_{data}(r)$].  Despite the fact that the deprojection amplifies
the fluctuations and errors present in the data, and although the
deprojection of the data introduces systematic errors due to the
effects of the {\tt WFPC2} PSF, at radii of $\simeq 0.1''$ (where the
effects of the {\tt WFPC2} PSF are modest) the comparison between
$\rho_{mod}(r)$ and $\rho_{data}(r)$ provides a rough estimate for the
effect of the central compact source on the underlying stellar
density, and more generally an estimate for the uncertainty associated
with the measurements.

In order to quantify the general accuracy of this estimate, and in
particular the effects of the {\tt WFPC2} PSF, we carried out the
following test. For the galaxies which do not host a central compact
source, we started from their (PSF-free) L95 models, deprojected both
these ``intrinsic'' profiles and those obtained from them after
convolution with the {\tt WFPC2} PSF, and compared the corresponding
densities at 0.1$''$. The deprojections of the intrinsic and
PSF-convolved profiles yielded densities at 0.1$''$ (or 10 pc) within
10-15\%. For the exponential galaxies which host a central compact
source, we added to their (PSF- and compact source- free) L95 models a
central source described by a Plummer model of varying radius, in
order to also assess the effect of such a light concentration on the
central densities. We then repeated the same steps described above,
i.e., deprojected these intrinsic profiles and those obtained after
PSF-convolution, and compared the resulting densities at 0.1$''$.  In
this case, the comparisons showed that for the range of compact source
luminosities and radii observed in our sample, the straight
deprojection of the data, ignoring the effect of the HST PSF, can be
in error by up to a factor 50\% (see Figure 2, where we plot, as an
example, the results for ESO482G17). This error is much smaller than
the effect of the compact source itself, which typically increases
$\rho_{0.1''}$, $\rho_{10pc}$ of a factor ten or more.

The deprojected luminosity density profiles are shown in Figure 3,
where the deprojections of the fit ($\rho_{mod}$) are plotted as solid
lines, and those from the data ($\rho_{data}$) as dashed lines.  When
two analytical descriptions were available for the same galaxy, in 
Figure 3 we plot the deprojection corresponding to the first of the two
fits in Table 2. The values of the stellar densities at $0.1''$ and at
10 pc derived from deprojecting the data ($\rho_{data,0.1''}$ and
$\rho_{data,10pc}$, respectively) and the models ($\rho_{mod,0.1''}$
and $\rho_{mod,10pc}$, respectively) are given in Table 3.

In the reminder of this paper, we use the values of
$\rho_{data,0.1''}$, $\rho_{mod,0.1''}$ and $\langle \gamma \rangle$
to describe the nuclear properties of disk galaxies.  The choice of
using the densities at $0.1''$ rather than those at 10 pc is justified
by the fact that the latter, although more physically motivated, often
probe a radius smaller than the typical FWHM of the PSF. Nonetheless,
we have checked that our main results remain unchanged when the
densities at $0.1''$ are substituded with those at 10 pc.  For those
objects for which two values of $\rho_{mod,0.1''}$ were available, we
adopted the smallest of the two. This choice maximizes the estimate
for the associated errorbar, since in these objects typically
$\rho_{data,0.1''} > \rho_{mod,0.1''}$.

\section{Discussion}

The HST exploration of the nuclei of nearby early-type galaxies has
highlighted the diversity between low-luminosity, rotation-supported,
radio-quiet, disky-isophote galaxies, and high-luminosity,
pressure--supported, radio-loud, boxy-isophote systems. The diversity
holds in fact down to the tens-of-parsecs scales: the former have
power-law (steep-cusp) light profiles and high stellar densities; the
latter show a break radius inside which the profile flattens to a
shallow (but non-zero) cusp slope, and have low-density nuclei (e.g.,
L95).  It is unclear however whether the connection between small and
large scale properties in early-type spheroidals is operated by
nature, i.e. it is the outcome of a different formation epoch, time
scale or physical mechanism, or is operated by nurture, i.e.  it
results from subsequent galactic evolution.  In the orthodox
framework where bulges of spiral galaxies are the extrapolation of the
spheroidal family toward fainter magnitudes, and in the light of the
variety of structural properties shown by the central spheroidal
structures embedded in disk galaxies (e.g., Kormendy 1993; Courteau et
al.\ 1996; papers I and II), the differences and similarities between
the nuclear properties of {\it (i)} $R^{1/4}$-law versus exponential
bulges, and {\it (ii)} bulges versus higher luminosity spheroidals,
can provide important feedback toward understanding the processes
involved in the formation of spheroidals and their nuclei.

In Figure 4 we plot the average nuclear slope inside 0.1$''$-0.5$''$,
$\langle \gamma \rangle$, versus the total F606W magnitude of the
spheroidal component $M_{F606W}$. Triangles represent galaxies fitted
by a single exponential, pentagons those fitted by two exponentials,
circles the $R^{1/4}$-law bulges, and asterisks the early-type
galaxies of L95, F97 (small symbols), and C97a, C97b (large symbols).
Only 33 of the 41 bulges for which we could derive the mesurements of
$\langle \gamma \rangle$ appear in the figure: for seven of the
remaining eight objects, no measurement of $M_{F606W}$ was available
(see Table 1, galaxies labelled by ``NGF'' in the last column --
ESO205G7 is not included in these seven object, since for it neither
$M_{F606W}$ nor $\langle \gamma \rangle$ could be derived);
furthermore, we excluded NGC 488 from the plots since its measurement
of $M_{F606W}$ is uncertain (although it definitely shows an
$R^{1/4}$-law profile, see Figures 6 and 8). We verified that the
inclusion of NGC 488 in our analysis does not change our conclusions.

There is a clear dichotomy between $R^{1/4}$-law and exponential
bulges, which holds even when only exponential bulges embedded in
otherwise normal outer disks (pentagons) are considered.  The
$R^{1/4}$-law bulges lie on the $\langle \gamma \rangle$ versus
absolute bulge magnitude relation traced by early-type galaxies of
similar luminosities, i.e. the fainter the bulge, the steeper its
nuclear stellar cusp. By contrast, exponential bulges (extend to
fainter magnitudes and) have significantly shallower stellar cusps for
a given luminosity. It is true that these typically underly the light
contributed by a central compact source, and are therefore subject to
a larger uncertainty; however, the dichotomy remains even within the
most conservative assumptions for the error bars. The effect is likely
not one of ``evolution'' along the Hubble sequence, since we compare
mostly galaxies of similar and intermediate, i.e. Sb to Sc, Hubble
type (see Figure 5, where we plot the average nuclear slope inside
0.1$''$-0.5$''$, $\langle \gamma \rangle$, versus Hubble type, from
RC3). In Figure 5, NGC 488 is included as a $R^{1/4}$-law bulge, while
NGC 3928, ESO548G29, ESO240G12 and ESO482G17 are excluded, due to
their uncertain morphological classification. The six squares in
Figure 5 (absent in Figure 4) correspond to those galaxies for which
no good fit to a bulge component could be derived, but for which the
measurement of $\langle \gamma \rangle$ could be performed (all
galaxies -- excluding NGC 3928 -- labelled with ``NGF'' in the last
column of Table 1).

A main question is whether the difference in nuclear slopes between
$R^{1/4}$-law and exponential bulges carries important information
about the structure (and formation) of these systems, or is rather a
trivial result which simply states the difference between an inward
extrapolation of an $R^{1/4}$-law and an exponential profile.  The
bulge parameters presented in papers I and II were obtained from fits
in the range 0.5$''$ to the last measured point: using the largest
radial range available was found to provide an overall better fit to
the data. However, performing the $R^{1/4}$-law and exponential fits
after excluding the data inside the innermost 1$''$ demonstrates that
there is no ambiguity outside such radius between $R^{1/4}$-law and
exponential bulges: the classification of a bulge as an $R^{1/4}$-law
or an exponential structure holds on a radial range entirely different
from the 0.1$''$-0.5$''$ used in the derivation of the average nuclear
slopes $\langle \gamma \rangle$ (see Appendix B for further details).
This strongly supports the interpretation that esponential bulges have
intrinsically shallower cusp slopes than $R^{1/4}$-law structures. In
Figure 6 we show the L95 fits (dotted lines), and the single bulge or
bulge plus disk fits (solid lines), for four representative objects,
i.e., a single exponential (E482G17) or $R^{1/4}$-law (NGC488) bulge,
a two-exponentials galaxy (E498G5), and a $R^{1/4}$-law plus
exponential object (NGC2344).  When a two-components fit is required,
we also plot the bulge (short-dashed lines) and disk (long-dashed
lines) components separately. The total (bulge plus disk) fits and the
bulge components overlap at small radii. These examples clearly
illustrate that while the L95 fits are aimed at representing the
innermost galactic regions, the bulge (plus disk) fits are aimed at
matching the outer profiles.

The deprojected stellar density at $0.1''$ versus absolute magnitude
of the bulge component $M_{F606W}$ is shown in Figure 7.  The symbols
(as in Figure 4) represent the measurements derived from the
deprojections of the analytical fits of Table 2 ($\rho_{mod,0.1''}$ in
Table 3). For those galaxies whose deprojection of the surface
brightness profile gives a stellar density $\rho_{data,0.1''}$ which
is significantly larger than $\rho_{mod,0.1''}$, an arrow is plotted,
which connects the value of $\rho_{mod,0.1''}$ to the corresponding
value of $\rho_{data,0.1''}$. The spherical deprojections might
underestimate the density for large intrinsic flattening. Assuming
that the exponential bulges have flattening comparable with that of
the $R^{1/4}$-law bulges, these structures have nuclear stellar
densities underlying the central compact sources which are at the
low-end side of those of early-type spheroidals and $R^{1/4}$-law
bulges. Exponential bulges might therefore break the general trend of
``classical'' spheroidals, i.e., to have a higher density the lower
the luminosity.  The ``dichotomy'' in nuclear properties between
exponential and $R^{1/4}$-law bulges constitutes an additional piece
of the puzzle concerning ``how and when'' bulges form. A discussion on
the possible implications of our results for scenarios of bulge
formation (generally supporting the idea that there is present-day
``bulge-formation'' in the centers of spirals through bar-driven
infall of dissipative material; e.g., Combes et al.\ 1990; Hasan,
Pfenniger \& Norman 1993; Norman, Sellwood \& Hasan 1996) is beyond
the scope of this paper, and is presented in Carollo (1998).

\section{Conclusions}

In this paper we have investigated the relation between the nuclear
structure of spiral galaxies and the physical properties of their
bulges. In particular, we have {\it (i)} modelled the {\tt WFPC2}
F606W nuclear surface brightness profiles of 41 spiral galaxies with
the analytical law introduced by L95 (data from papers I and II), and
{\it (ii)} deconvolved the surface brightness profiles and their
analytical fits in order to estimate the nuclear stellar densities of
disk galaxies.

Our main result is that $R^{1/4}$-law bulges and exponential bulges
have significantly different nuclear stellar cusps and densities.
Specifically, $R^{1/4}$-law bulges have steep stellar cusps which
steepen with decreasing luminosity; furthermore, their stellar cusp
slopes and densities are similar in values to those of early-type
systems of comparable luminosity. By contrast, in exponential bulges,
the inward extrapolations underlying the light from the compact
sources which sit in their very centers imply rather shallow stellar
cusps and, very likely, relatively low nuclear stellar densities.

\bigskip
\bigskip

\noindent
{\bf Acknowledgements}

We heartly thank Tim Heckman and Colin Norman for helpful discussions,
and the anonymous referee for constructive comments to an earlier
version of this paper. CMC is supported by NASA through the grant
HF-1079.01-96a awarded by the Space Telescope Institute, which is
operated by the Association of Universities for Research in Astronomy,
Inc., for NASA under contract NAS 5-26555. This research has been partially
funded by grant GO-06359.01-95A awarded by STScI, and has made use
of the NASA/IPAC Extragalactic Database (NED) which is operated by the
Jet Propulsion Laboratory, Caltech, under contract with NASA.

\bigskip
\bigskip

\noindent
\begin{center}{{\bf Appendix A.} Details on the Analytical Fits}\end{center}

\medskip

We used the same fitting procedure described in C97a; we addess to
this reference for further information. This was carried out in two
steps: (1) The first step isolated primary from secondary minima of
$\chi^2$. The $\chi^2$ values were computed on a grid of points
uniformly distributed on a wide hypercube in parameter space (with
dimension equal to the number of free parameters). Once a minimum
value was found, a new, smaller, hypercube was placed on that
location, and the procedure iterated. (2) The minimum value found on
the hypercube was then used as starting point to initialize a downhill
simplex minimization. We tested the procedure on simulated data, and
verified that it recovered the initial values with high accuracy.  We
accepted as final the fits associated with the absolute minimum of
$\chi^2$.

Deconvolutions of {\tt WF/PC} data have been proven to be very
reliable (e.g., L95). However, in our analysis, we chose not to apply
any deconvolution to the post-refurbishment {\tt WFPC2} images, and to
correct for PSF-blurring while modeling the light profiles.
Therefore, the models were convolved with the appropriate PSFs before
being compared to the data.  Since pointlike sources with adequate S/N
located near to the nuclei were not available for most of the
galaxies, we computed the PSFs by running Tinytim (Krist 1992).  Focus
drifts and breathing modify the PSF profile and affect the flux within
a 1 PC pixel radius up to 10\% (and within 5 PC pixels up to 5\%;
Suchkov \& Casertano 1997).  Therefore, the simulated PSFs obtained by
construction at the nominal focus position are in principle of similar
quality than PSFs derived from archival stars.  Furthermore, our
approach of convolving the models rather than deconvolving the data
minimizes the effects of using a possibly non-perfect PSF.

\bigskip
\bigskip

\noindent
\begin{center}{{\bf Appendix B.} Classifying $R^{1/4}$-law or exponential bulges outside $1''$}\end{center}

\medskip

In order to ensure that the classification of a bulge as an
$R^{1/4}$-law or an exponential structure is valid on a radial range
entirely different from that used in the derivation of $\langle \gamma
\rangle$ (equal to 0.1$''$-0.5$''$), we performed as a test the
$R^{1/4}$-law and exponential fits after excluding the data inside the
innermost $1''$.  This value is a compromise between a radius large
enough to exclude the range where $\langle \gamma \rangle$ is
computed, and small enough to still allow the detection of the small,
disk-embedded exponential bulges (pentagons in the figures).  As an
example, the results of the test are illustrated in Figure 8 for the
same four galaxies presented in Figure 6. The solid lines represent
either a single exponential (left panels) or a double exponential
(right panels) fit; the dashed lines represent either a single
$R^{1/4}$-law (left panels) or an $R^{1/4}$-law plus exponential
(right panels) fit. Two different scales are used for the abscissa for
the galaxies in the left and right panels, consistently with the
different scales of their bulge components. An offset of two
magnitudes has been applied to ESO482G17 for plotting purposes. There
are two important points to note: {\it (i)} the kind of profile which
provides the bulge classification given in papers I and II,
i.e. exponential or $R^{1/4}$-law, still provides a better fit to the
inner galactic regions, even when the innermost $1''$ is excluded from
the fits; {\it (ii)} the alternative profile with respect to the one
that provides the classification given in papers I and II generally
provides (not only a worse fit but also) physically meaningless best
fit parameters (e.g., for ESO482G17, the $R^{1/4}$-law best fit of
Figure 8 has an $R_e\sim385''$).  We conclude that the distinction
between $R^{1/4}$-law and exponential bulges holds in a radial range
which excludes the one used to derive $\langle \gamma \rangle$, and
that the difference in nuclear cusp slopes $\langle \gamma \rangle$
between $R^{1/4}$-law and exponential bulges has a physical origin.
We retained the bulge parameters presented in papers I and II for our
discussion, since those fits gave an overall better description of the
profiles.

\newpage

{\bf Figure 1.} The surface brightness profiles, and the fits with the
PSF-convolved analytical models (for the galaxies for which they could
be derived, solid lines).
   
\bigskip

{\bf Figure 2.} The effect of the {\tt WFPC2} PSF on the deprojected
density at 0.1$''$.  For ESO482G17, we plot the relative density error
($\rho_{NO-PSF}$-$\rho_{PSF}$)/$\rho_{PSF}$ versus the radius of a
simulated compact source. $\rho_{NO-PSF}$ is the density derived by
deprojecting the L95 model added of the Plummer-modeled compact
source, i.e., is the PSF-free measurement. $\rho_{PSF}$ is the density
derived by deprojecting the profile obtained from the previous one
after convolution with the {\tt WFPC2} PSF.

\bigskip

{\bf Figure 3.} Stellar density profiles obtained by deprojecting the
observed surface brightness profiles (dashed lines) and their best
fits (solid lines).
   
\bigskip

{\bf Figure 4.} Average nuclear slope $\langle \gamma \rangle$ inside
0.1$''$-0.5$''$ versus absolute F606W magnitude of the spheroidal
component. Triangles are the single exponential galaxies, pentagons
the double exponential bulges, circles the $R^{1/4}$-law bulges, and
asterisks the early-type galaxies of L95 and F97 (small symbols) and
C97a, C97b (large symbols).  Conservative error bars are shown in the
upper left corner of the diagram for the galaxies which host a central
compact source (large), and for the remaining galaxies (small).

\bigskip
 
{\bf Figure 5.} Average nuclear slope inside 0.1$''$-0.5$''$ ($\langle
\gamma \rangle$) versus Hubble type (from RC3). Triangles are the
single exponential galaxies, pentagons the double exponential bulges,
circles the $R^{1/4}$-law bulges, and squares the galaxies for which
no bulge-component could be fitted.

\bigskip

{\bf Figure 6.} The L95 fit (dotted line), and the single bulge or
bulge plus disk fits (solid line), for four representative galaxies.
E482G17 is fitted with a single exponential, NGC488 with a single
$R^{1/4}$-law.  For the two-components falaxies (E498G5,
two-exponentials, and NGC2344, $R^{1/4}$-law plus exponential), we
plot the bulge (short-dashed line) and disk (long-dashed line)
components separately. The total (bulge plus disk) fits and the bulge
component overlap at small radii.

\bigskip

{\bf Figure 7.} Stellar density at $0.1''$ versus absolute F606W
magnitude of the bulge component.  Symbols are as in Figure 3.  For
our sample, they represent the measurements relative to the
deprojections of the analytical fits to the surface brightness
profiles ($\rho_{mod,0.1''}$). For the galaxies for which the stellar
density at $0.1''$ obtained from the deprojection of the observed
surface brightness profile ($\rho_{data,0.1''}$) is significantly
larger than $\rho_{mod,0.1''}$, an arrow is plotted, which connects
the value of $\rho_{mod,0.1''}$ to the corresponding value of
$\rho_{data,0.1''}$.
   
\bigskip

{\bf Figure 8.} Exponential and $R^{1/4}$-law fits performed as a test
outside $1''$.  The results for the same four galaxies of Figure 6 are
presented. The solid lines are either a single exponential (left
panels) or a double exponential (right panels) fit; the dashed lines
are either a single $R^{1/4}$-law (left panels) or an $R^{1/4}$-law
plus exponential (right panels) fit.  An offset of two magnitudes has
been applied to ESO482G17. The kind of profile which provides the
bulge classification given in papers I and II still provides a better
fit to the inner profiles, even when the innermost $1''$ is excluded
from the fits.
   
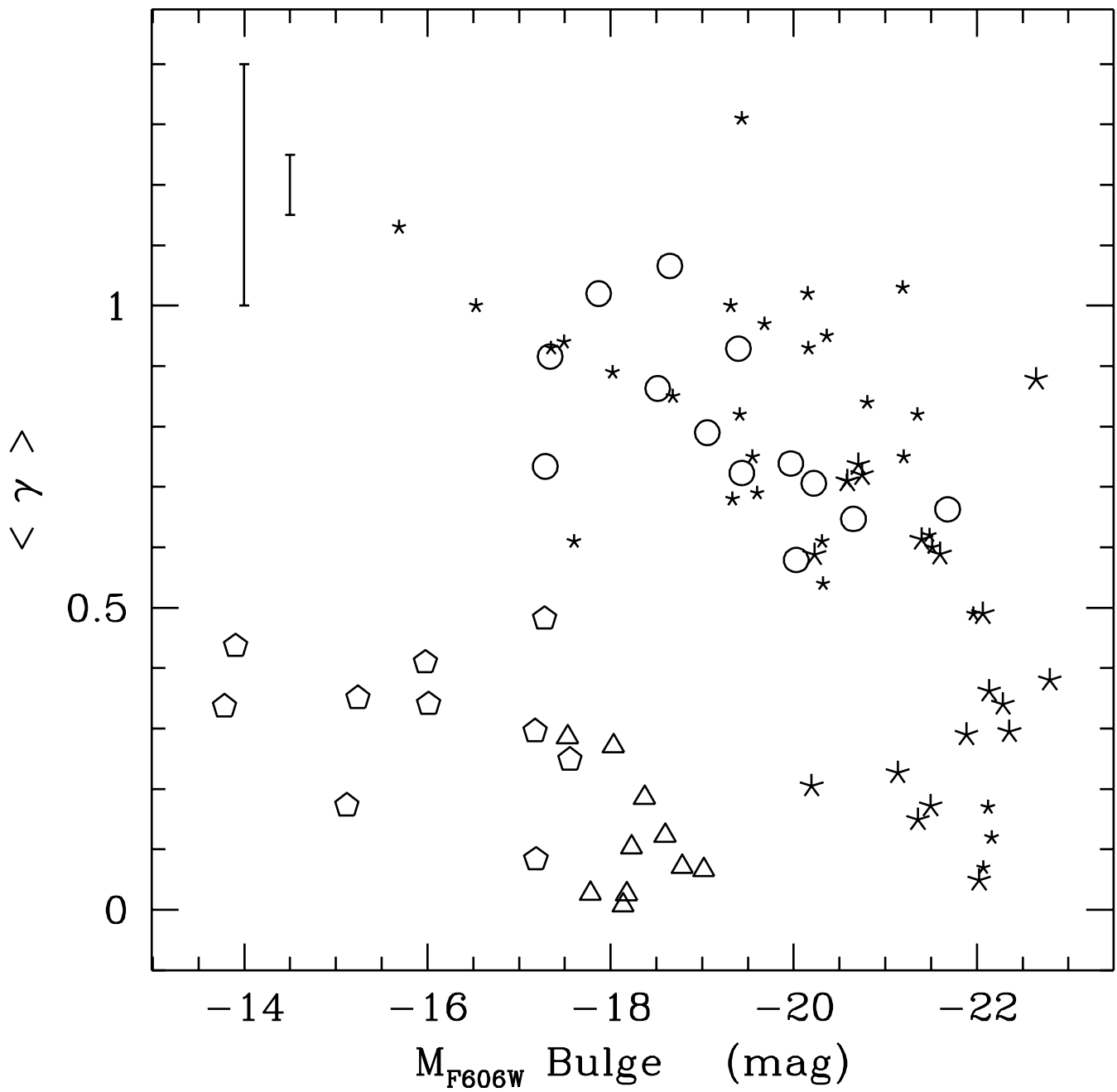
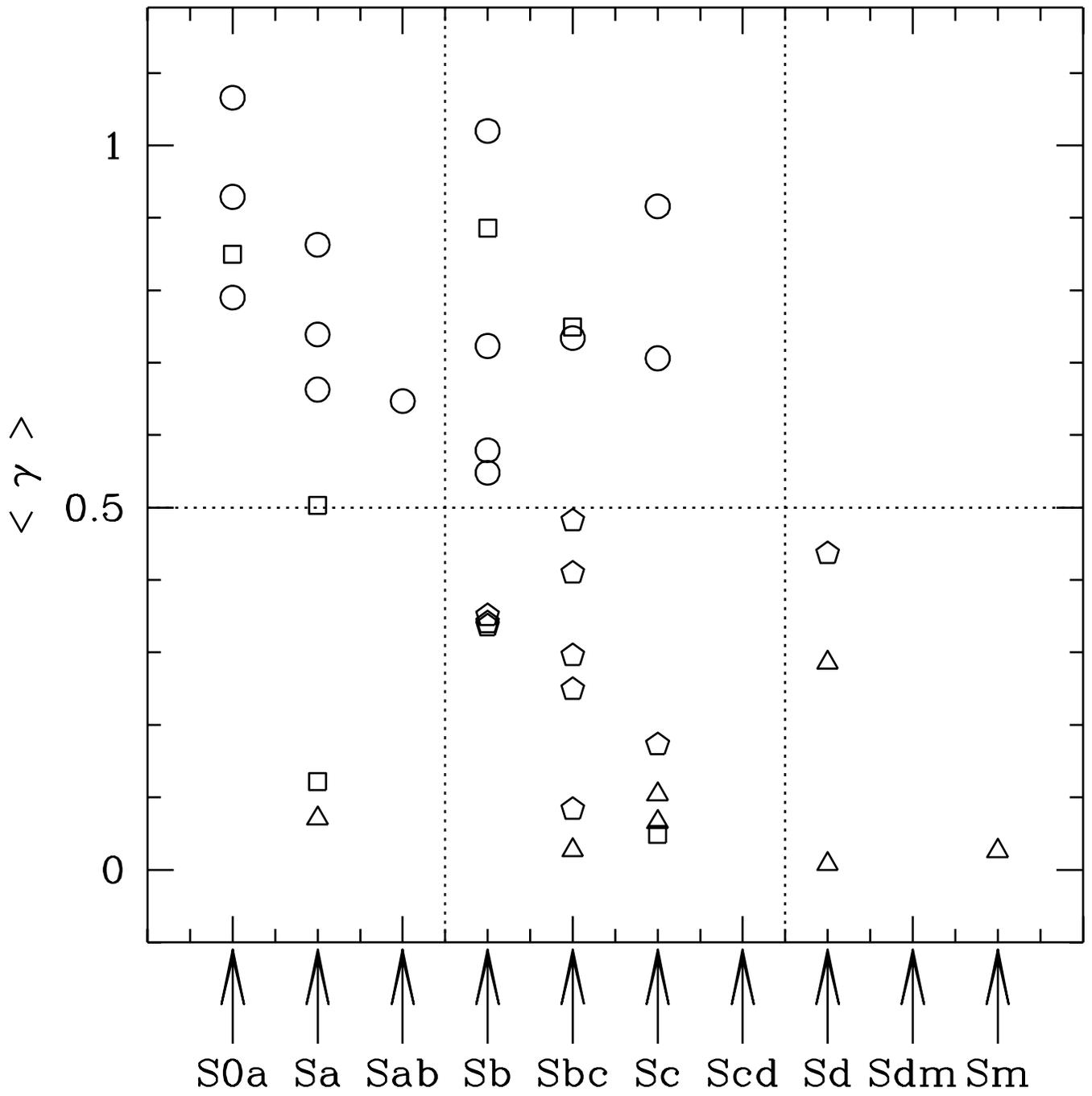
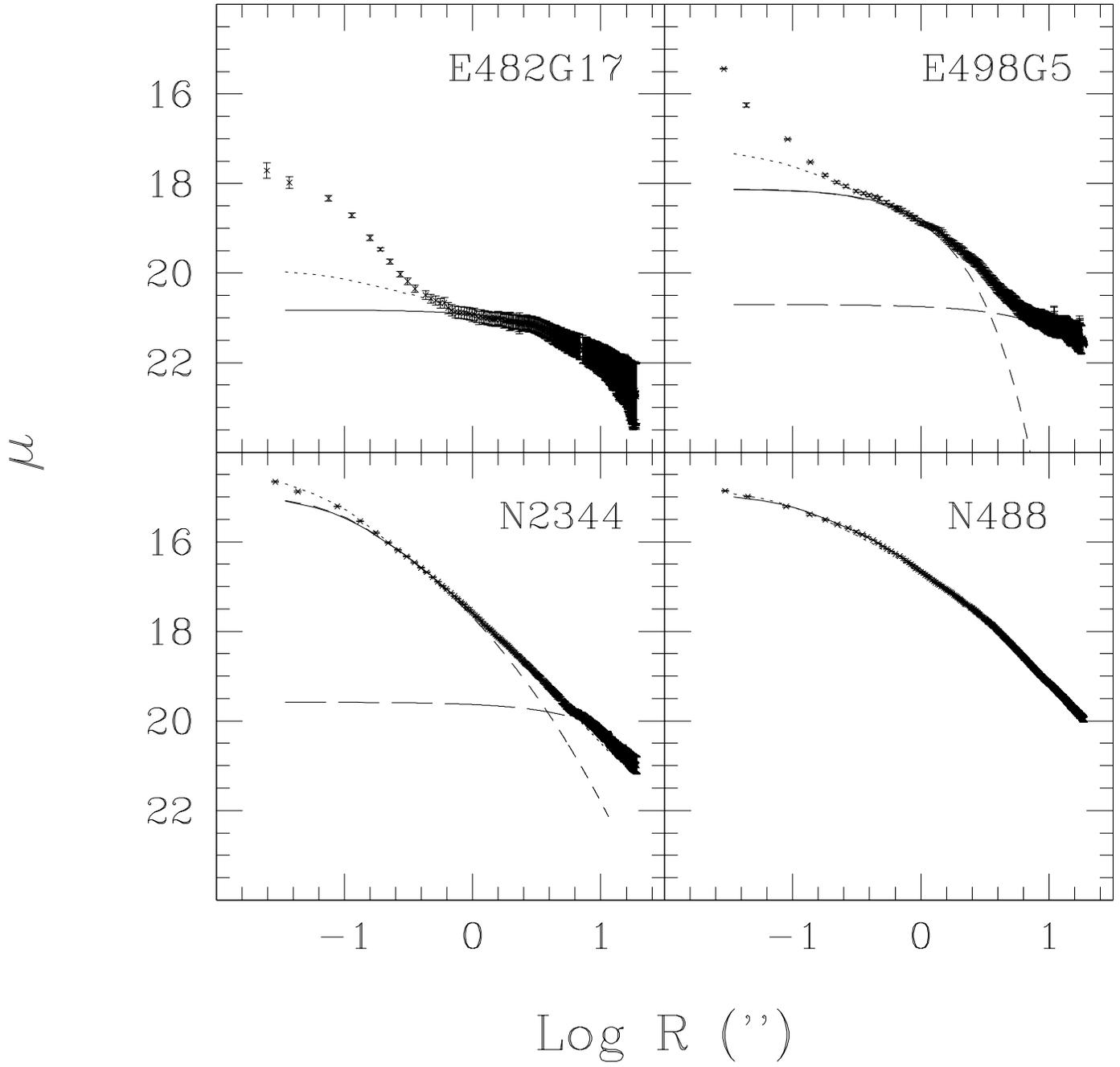
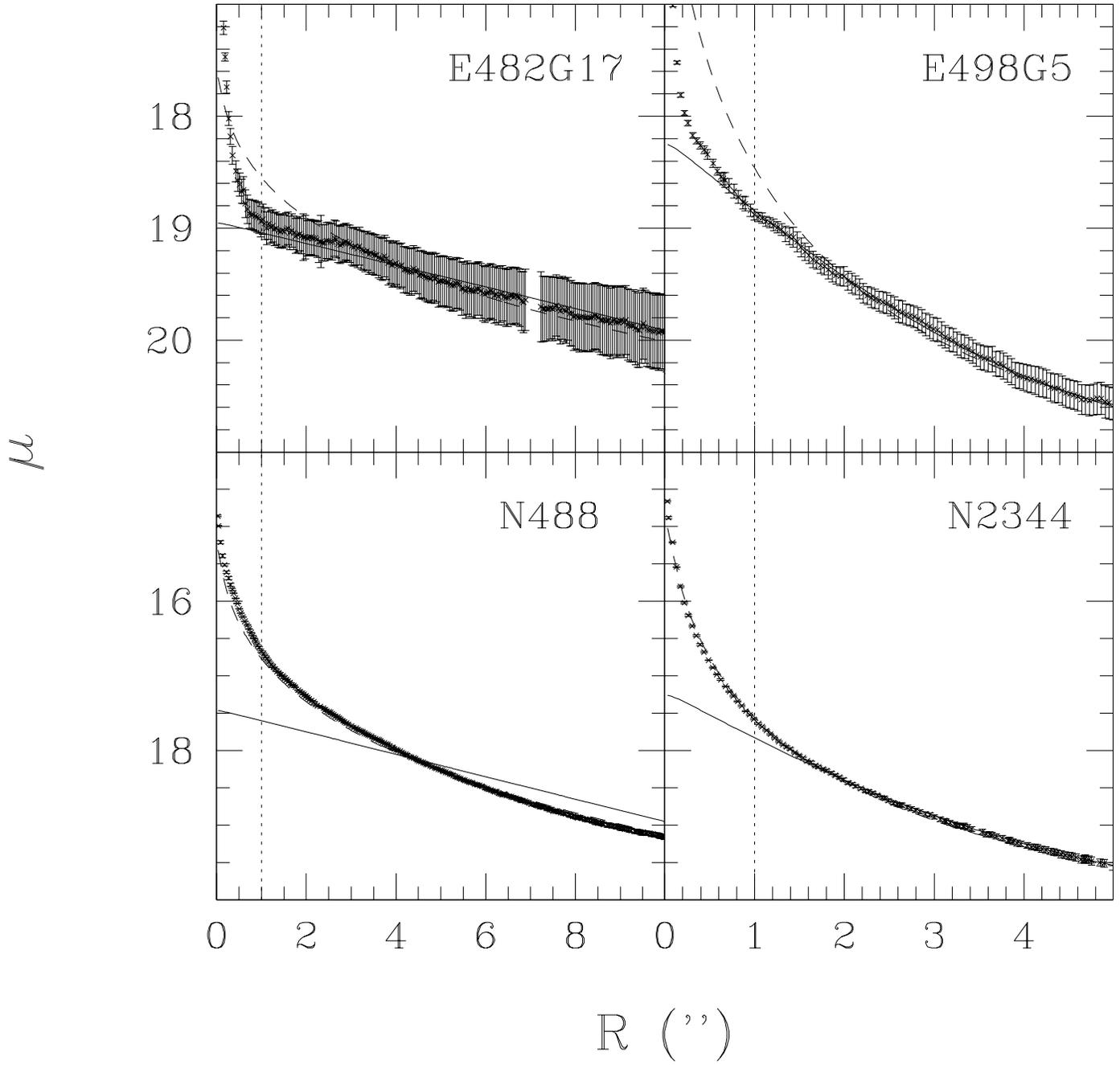
\end{document}


\begin{table*}
{\tiny\begin{center}\begin{tabular}{lcccccc}
\hline\hline
\multicolumn{1}{l}{Name} &
\multicolumn{1}{c}{$D$} &
\multicolumn{1}{c}{$B$} &
\multicolumn{1}{c}{Type} &
\multicolumn{1}{c}{Compact Source$^\clubsuit$} &
\multicolumn{1}{c}{Bar$^\heartsuit$} &
\multicolumn{1}{c}{Bulge$^\spadesuit$} \\
\multicolumn{1}{l}{} &
\multicolumn{1}{l}{(Mpc)} &
\multicolumn{1}{l}{(mag)} &
\multicolumn{1}{l}{} &
\multicolumn{1}{c}{} &
\multicolumn{1}{c}{} &
\multicolumn{1}{c}{(mag, $''$)} \\
\hline 
ESO 205 G7              &31 &15.02 &.SBR3../Sb      & Y  & Y& NGF  \\
ESO 240 G12$^{expo1}$     &28 &14.35 & .S?..../Sa-b & Y$^1$  & ?& 13.64, 9.67       \\
ESO 317 G20$^{R^{1/4}}$   &38 &12.66 &.SAT5*./Sa    & N  & N& 12.68, 11.83   \\
ESO 482 G17$^{expo1}$     &24 &15.10 &.L?..../Sa-b  & Y  & ?& 13.87, 16.79     \\
ESO 498 G5$^{expo2}$&37 &13.81 &.SXS4P./Sb          & Y  & ?& 15.56, 2.18      \\
ESO 499 G37$^{expo1}$      &14 &13.15 &.SXS7*./Sb   & Y  & ?& 13.20, 17.52     \\
ESO 508 G34$^{expo1}$     &30 &14.50 &.SXS9*./Sb    & Y  & ?& 14.21, 9.97    \\
ESO 548 G10$^{expo2}$     &20 &13.30 &.SXT7*./Sb-c  & Y  & ?& 17.60, 2.33      \\
ESO 548 G29$^{expo1}$     &24 &14.19 & SB?/Sb       & Y  & Y& 13.53, 12.13     \\
ESO 549 G18$^{expo2}$     &24 &13.49 & .SXT5../Sb   & Y  & Y& 16.78, 3.75   \\
ESO 572 G22$^{expo1}$     &27 &14.92 & RSB.7P?/Sb   & Y  & Y& 14.02, 10.41  \\
NGC 406$^{expo1}$       &23 &12.88& .SAS5*./Sb      & Y  & ?& 12.79, 13.07   \\
NGC 488$^{R^{1/4}}$          &37 &11.15& .SAR3../S  & N  & N& 10.2, 40.0 (?)     \\
NGC 1325$^{expo2}$           &27 &12.30&.SAS4../Sd  & Y  & N& 16.18, 1.85   \\
NGC 1345$^{expo1}$        &24 &13.57& SBS5P*/SBa    & Y  & Y& 13.67, 5.89      \\
NGC 1483$^{expo1}$        &16 &13.23&SBS4*./Sb-c    & Y$^1$  & Y& 13.24, 13.33     \\
NGC 1640$^{R^{1/4}}$       &25 &12.48&.SBR3../S(r)a & N  & ETB& 14.12, 2.47   \\
NGC 2082$^{expo2}$      &19 &12.64&.SBR3../Sb       & Y  & Y& 15.38, 3.70   \\
NGC 2196$^{R^{1/4}}$         &35 &11.66&PSAS1../Sa  & N  & N& 11.04, 22.94  \\
NGC 2344$^{R^{1/4}}$          &14 &12.68& SAT5*./Sb & N  & N& 13.39, 6.18      \\
NGC 2460$^{R^{1/4}}$         &26 &12.38& .SAS1../Sb & N  & N& 13.56, 3.93      \\
NGC 2758$^{expo2}$        &30 &13.46& PSB.4P?/Sb:   & Y  & ?& 15.20, 2.81      \\
NGC 3045$^{expo2}$           &35 &13.61& .SAR3?./Sb & N  & ?& 17.48, 0.86   \\
NGC 3054$^{R^{1/4}}$          &34 &11.95& .SXR3../Sb& N  & ETB& 12.63, 9.31   \\
NGC 3177$^{expo2}$         &19 &13.07&.SAT3../Sb    & Y  & N& 15.58, 0.42   \\
NGC 3259$^{R^{1/4}}$      &27 &12.97& .SXT4*./Sb    & Y$^1$  & N& 14.87, 7.44      \\
NGC 3277                  &22 &12.55&.SAR2../Sa-Sb  & N  & N& NGF           \\
NGC 3455$^{expo2}$     &19 &12.87&PSXT3../Sb        & Y  & ?& 17.61, 1.53  \\
NGC 3898$^{R^{1/4}}$         &22 &11.63&SAS2../Sa   & N  & N& 11.06, 11.74     \\
NGC 3900$^{R^{1/4}}$         &27 &12.30&.LAR+../Sa  & Y  & N& 13.50, 2.85      \\
NGC 3928                     &17 &12.94&.E?..../-   & Y  & N & NGF                 \\
NGC 4030$^{expo2}$            &20 &11.37& .SAS4../Sb& Y  & N & 14.33, 2.0    \\
NGC 4260                     &29 &12.61& .SBS1../SBa& N  &ETB & NGF            \\
NGC 4384                     &37 &13.54& .S..1../Sa & Y  &? &  NGF                \\
NGC 4806                    &37 &13.09&.SBS5?./Sb   & Y  & N & NGF            \\
NGC 4980$^{expo1}$        &22 &13.19&.SXT1P?/Sa     & Y  & ?& 12.93, 12.81  \\
NGC 5121$^{R^{1/4}}$          &23 &11.27&PSAS1../Sa & N  & N& 11.84, 7.10  \\
NGC 5985$^{R^{1/4}}$          &46 &11.67&.SXR3../Sb & N  & ETB?& 13.88, 8.0   \\
NGC 6340$^{R^{1/4}}$        &18 &11.82& .SAS0../Sa  & N  & N& 12.22, 8.03      \\
NGC 6384$^{expo2}$           &26 &11.29&.SXR4../Sb  & Y  & Y& 14.52, 1.95   \\
NGC 7280$^{R^{1/4}}$      &28 &13.00&.LXR+../S0-Sa  & N  & N& 12.84, 3.51      \\
NGC 7421                    &26 &12.62&.SBT4../Sb-c & Y  & Y& NGF          \\
NGC 7690                    & 21&13.03& .SAR3*\$/Sb & Y  & N& NGF \\
\hline
\end{tabular}\end{center}}
\caption{ Parameters for the 43 galaxies. Distance $D$ and total $B$
magnitude are from the RC3 catalogue (de Vaucouleurs et al.\
1991). Morphological type is from the ESOLV (Lauberts \& Valentijn
1989) or UGC (Nilson 1971) catalogues (left), or from the RC3
catalogue (right).  The information provided in the other columns is
extracted from papers I and II.  $\clubsuit$ A ``Y'' indicates the
presence of a central compact source, a ``N'' stands for no central
compact source. $^1$ The central source in ESO 240 G12, NGC 1483 and
NGC 3259 is unresolved. $\heartsuit$ A ``Y'' indicates the presence of
a bar, a ``N'' stands for no barred component.  A question mark
indicates that a faint bar might be present, or that the entire galaxy
has a flattened, bar-like (irregular) morphology. ``ETB'' indicates
the presence of an early-type bar.  $\spadesuit$ The column lists the
total apparent magnitude in F606W and the half-light radius $R_e$ (in
$''$) of the bulges. Close to the galaxy name, the superscript
$R^{1/4}$, {\it expo1} or {\it expo2} identifies galaxies whose
bulge-like component has been fitted with a $R^{1/4}$-law or an
(single or double) exponential profile. All magnitudes are corrected
for Galactic extinction using the values published by Burstein \&
Heyles (1984).}
\end{table*}
 
\normalsize

\newpage

\begin{table*}
{\tiny\begin{center}\begin{tabular}{lccccccc}
\hline\hline
\multicolumn{1}{l}{Name} &
\multicolumn{1}{c}{Range} &
\multicolumn{1}{c}{$\alpha$} &
\multicolumn{1}{c}{$\beta$} &
\multicolumn{1}{c}{$\gamma$} &
\multicolumn{1}{c}{$R_b$} &
\multicolumn{1}{c}{$\mu_b$} &
\multicolumn{1}{c}{$\langle \gamma \rangle$} \\
\multicolumn{1}{l}{} &
\multicolumn{1}{c}{$('')$} &
\multicolumn{1}{c}{} &
\multicolumn{1}{c}{} &
\multicolumn{1}{c}{} &
\multicolumn{1}{c}{($''$)} &
\multicolumn{1}{c}{(mag)} &
\multicolumn{1}{c}{} \\
\hline 
ESO 205 G7    &-&-&-&-&-&-&- \\
ESO 240 G12   &0.16-10&1.362 &1.799 &0.182&6.506 & 20.70&  0.188 \\
              &0.7-10&1.147&1.560&0.000&4.190   &20.29 & 0.058 \\
ESO 317 G20   &0-10&4.850&0.937&0.498&0.220   &16.17 & 0.706  \\
ESO 482 G17   &0.3-10&70.40&0.596&0.290&3.669   & 21.26&  0.271   \\
ESO 498 G5    &0.16-10&5.582&1.007&0.522&1.052   & 18.87&   0.482  \\
ESO 499 G37   &0.7-10&5.548&0.823&0.306&5.101   & 20.80&  0.286  \\
ESO 508 G34   &0-10&1.497&1.599&0.000&4.021   &20.85 &  0.026 \\
ESO 548 G10   &0.3-10&19.22&0.704&0.459&3.548   & 20.82&  0.437    \\
ESO 548 G29   &$^*$&3.0&0.90&0.20& 3.50  & 20.86 & 0.186 \\
ESO 549 G18   &0.16-10&5.167&0.682&0.360&3.666   & 20.86& 0.339  \\
              &0.7-10&1.231&0.819&0.008&2.153   &20.59 & 0.006\\
ESO 572 G22   &0.16-10&1.945&1.413&0.000&3.763   &20.62 & 0.008  \\
NGC 406       &0.2-10&3.963&1.315&0.072&4.645   & 19.82& 0.066 \\
NGC 488       &0-10&0.643&2.148&0.419&8.817   &19.01 & 0.548 \\
NGC 1325      &0.16-10&0.302&0.975&0.208&1.050   & 18.88& 0.480 \\
              &0.25-10&3.359&0.674&0.343&0.690   &18.59 & 0.340\\
NGC 1345      &0.3-10&6.973&1.454&0.113&2.026   &18.83 & 0.104 \\
NGC 1483      &0.3-10&1.365&1.060&0.000&3.724   &20.28 & 0.027  \\
NGC 1640      &0-10&59.85&1.024&0.895&0.020   &13.05 & 1.020  \\
NGC 2082      &0.16-10&20.74&0.652&0.345&2.511   & 19.42& 0.324\\
              &0.7-10& 49.17&0.654& 0.378&2.680&19.50 & 0.357 \\
NGC 2196      &$^*$&2.979&1.195&0.695&1.702   &17.21 & 0.663 \\
NGC 2344      &0-5&0.415&1.435&0.694&1.177   & 17.77& 0.916  \\
NGC 2460      &0-10&5.180&1.726&0.926&8.300   &19.26 & 0.913   \\
              &0.2-7&1.942&0.990&0.827&1.042   &17.03 & 0.813\\
NGC 2758      &0.3-10&1.756&1.318&0.000&1.515   &18.78 & 0.060\\
              &0.16-10&1.319&1.452&0.000&1.703   &18.88 & 0.107 \\
NGC 3045      &0.11-5&70.34&0.590&0.370&0.551   & 18.51& 0.351\\
NGC 3054      &0-10&6.382&1.179&0.553&0.441   &16.26 & 0.579 \\
NGC 3177      &-&-&-&-&-&-&- \\
NGC 3259      &0.3-5&0.156&0.441&1.026&0.120   & 17.07& 0.690  \\
              &0.3-10&0.214&0.040&1.292&1.445   &18.98 & 0.777\\
NGC 3277      &0-10&54.61&1.140&0.869&0.594   &16.21 & 0.850   \\
NGC 3455      &0.3-10&0.443&0.621&0.299&3.583   & 20.08& 0.351 \\
              &0.2-5&0.536&0.734&0.251&3.422   & 20.07& 0.323\\
NGC 3898      &0-10&0.916&1.499&0.304&0.553   & 15.29& 0.647 \\
NGC 3900      &0-10&0.123&1.613&0.600&0.565   & 16.28& 1.078  \\
              &0.2-10&0.099&2.011&0.252&1.535   &17.43 & 1.054\\
NGC 3928      &0.3-10&87.84&1.767&0.657&2.431   &17.98 & 0.620 \\
NGC 4030      &0.3-10&93.68&0.759&0.316&1.059   & 16.98& 0.296\\
NGC 4260      &0-10&0.365&2.426&0.000&7.464   &19.57 & 0.503 \\
NGC 4384      &0-10&1.960&1.099&0.000&0.764   &18.11 & 0.122  \\
NGC 4806      &$^*$&1.673&0.629&0.000&1.138   &19.36 & 0.049\\
NGC 4980      &0-10&379.0&0.669&0.077&2.965   &19.67 & 0.071\\
NGC 5121      &0-10&56.22&1.408&.799&0.886    &15.88 & 0.772\\
	      &$^*$&13.52&1.402&0.738&0.819   &15.86 &0.706\\
NGC 5985      &0.1-10&41.85&1.046&0.754&2.346   &18.78 &0.723\\
NGC 6340      &0-10&1.728&1.238&0.716&0.538   &15.98 &  0.790  \\
NGC 6384      &0.3-10&0.843&1.696&0.000&1.803   &17.78 & 0.249 \\
NGC 7280      &0.2-10&73.65&1.472&0.914&0.772   &16.16 &   0.900   \\
              &0-10&15.42&1.472& 0.965&0.802  &16.24 & 0.958\\
NGC 7421      &0.3-10&1.677&1.098&0.743& 2.720  &19.02 & 0.718 \\
              &0.2-10&9.315&1.012&0.806&2.490   &18.92 &0.779 \\
NGC 7690      &0-10&144.8&1.965&0.902&4.534   &18.57 & 0.886\\
\hline
\end{tabular}\end{center}}
\caption{The best fit values for $R_b$, $\alpha$, $\beta$, $\gamma$,
$\mu_b=-2.5 \log I_b$ and $\langle \gamma \rangle$. The column
``Range'' gives the radial interval over which the fits have been
performed. When two fits are reported for the same galaxy, the upper
values correspond to the dashed lines in Figure 1, and the lower
values correspond to the solid lines. The average mean of the two
corresponding values of $\langle \gamma \rangle$ is used in the
analysis. $^*$ indicates that no formal fit was performed, and the
corresponding parameters optimized by visual inspection.}
\end{table*}
 
\normalsize

\newpage

\begin{table*}
{\tiny\begin{center}\begin{tabular}{lcccc}
\hline\hline
\multicolumn{1}{l}{Name} &
\multicolumn{1}{c}{$\rho_{data,0.1''}$} &
\multicolumn{1}{c}{$\rho_{data,10pc}$} &
\multicolumn{1}{c}{$\rho_{mod,0.1''}$} &
\multicolumn{1}{c}{$\rho_{mod,10pc}$} \\
\multicolumn{1}{c}{} &
\multicolumn{1}{c}{($L_\odot pc^{-3}$)} &
\multicolumn{1}{c}{($L_\odot pc^{-3}$)} &
\multicolumn{1}{c}{($L_\odot pc^{-3}$)} &
\multicolumn{1}{c}{($L_\odot pc^{-3}$)} \\
\hline 
E240g12    &   27   &   59     &   5   &   8     \\
           &        &          &   1    &  1       \\
E317g20    &   172   &   211     &    198  &    487     \\
E482g17    &   40  &  52      &   3    &   4    \\
E498g5    &   134   &   813    &   37   &   93    \\
E499g37    &   63   &   43     &  10  &   7     \\
E508g34    &  0.7   &   1     &  0.4  &  0.4     \\
E548g10    &   198   &   198     &   14   &   14     \\
E548g29    &   42   &   64     &   4   &   5   \\
E549g18    &   62   &   91     &   7   &   9     \\
           &        &          &   0.7   & 0.8       \\
E572g22    &   40   &   74    &  0.2   &  0.2     \\
N1325    &      173&      344   &      44&      66     \\
           &        &          &   27    &     40    \\
N1345    &      41&      80   &      8&      9    \\
N1483    &      6&      6   &     0.9&     0.9     \\
N1640    &      1027&      1431   &     995      &      1568     \\
N2082    &      110&      93  &      26&      24     \\
           &        &          &     30  &    28     \\
N2196    &      498&      900   &      491&      1272    \\
N2344    &      1156&      631  &      1207&      623    \\
N2460    &      981&      1631   &    1016  &      1689     \\
           &        &          &    807   &    1305     \\
N2758    &      79&      193   &      3&      3    \\
           &        &          &     6  &      6   \\
N3045    &      31&      87   &      22&      46    \\
N3054    &      318&      423   &      317&      713     \\
N3259    &      251&      495   &      107&      182     \\
           &        &          &  131     &   232      \\
N3277    &      1559&      1849   &     1354 &      1620     \\
N3455    &      267&      221   &      20&      19     \\
           &        &          &    18   &      17   \\
N3898    &      1311&      1452   &     1578 &      1784     \\
N3900    &      1481&      2531   &     1629 &      3006     \\
           &        &          &     1586  &     2883    \\
N3928    &      3900&      2199   &   486   &      371    \\
N4030    &      861&      861  &      151   &   151     \\
N406     &      3&      4   &      2&      2     \\
N4260    &      84&      191   &      176&      272     \\
N4384    &      6&     -   &      7&      7     \\
N4806    &      63&      324   &     1&      1     \\
N488     &      301&      322   &     328  &       783     \\
N4980    &   1   &      2   &   2  &      2     \\
N5121    &  1704      &  2112      &  1647    &  2097         \\
           &          &          &   2005    &  2575       \\
N5985    &   129   &       485   &   125   &      542     \\
N6340    &   1425   &      1192   &   1593   &      1335     \\
N6384    &   224   &      350  &   62   &      71     \\
N7280    &   1829   &      2906   &   1594  &      3041     \\
           &        &          &   1800    &   3496      \\
N7421    &   513   &      859   &   218   &      346     \\
           &        &          &  256     &    412     \\
N7690    &   1184   &      1315   &   1096   &      1205     \\
\hline
\end{tabular}\end{center}}
\caption{Stellar densities at $0.1''$ and 10 pc obtained from
deprojecting the surface brightness profiles (subscript $data$) and
the best fitting models of Table 2 (subscript $mod$). For the galaxies
for which two best fits are available, the corresponding values for
$\rho_{mod,0.1''}$ and $\rho_{mod,10pc}$ are reported following the
same order as in Table 2.}
\end{table*}
 
\normalsize